\documentclass[11pt]{article}

\usepackage[utf8]{inputenc}
\usepackage[letterpaper,margin=1in]{geometry}
\usepackage{amsmath}
\usepackage{booktabs}
\usepackage{array}
\usepackage{microtype}
\usepackage[round,authoryear]{natbib}
\usepackage[hidelinks]{hyperref}
\usepackage{xurl}
\usepackage[font=small,labelfont=bf]{caption}
\usepackage{listings}
\usepackage{longtable}
\newcommand{\fld}[1]{\textsf{#1}}          
\newcommand{\var}[1]{\texttt{#1}}          
\newcommand{\defn}[1]{\textit{#1}}         

\hypersetup{
  pdftitle={Where ADS Crashes Differ: A Harmonized Comparison Against Human Baselines},
  pdfauthor={Doyeon Jang},
  pdfkeywords={automated driving systems, crash composition, NHTSA Standing General Order, CRSS, quasi-induced exposure, large language models}
}

\title{Where ADS Crashes Differ:\\ A Harmonized Comparison Against Human Baselines}
\author{Doyeon Jang\\ \small\texttt{doyeonjang0830@gmail.com}}
\date{November 2026}

\begin{document}
\maketitle

\begin{abstract}
Crash-rate comparisons between automated driving systems (ADS) and human drivers require an exposure denominator that only a few operators publish. We instead ask a composition question: given a crash, does the mix of crash types differ between ADS and human-driven vehicles? The question needs no mileage and can therefore cover every ADS operator. We assemble the NHTSA Standing General Order (SGO) ADS incident files across both file generations into 2,450 distinct in-scope incidents (2021--2026), of which 828 fall within the 2021--2024 window covered by the Crash Report Sampling System (CRSS), the national police-reported human comparator. A schema-constrained large language model (LLM) codes each SGO narrative into the CRSS crash-type taxonomy. Because SGO compels reporting far below the police-reporting threshold, we bracket reportability with an objective severity floor and a reasonable-person rubric, and claim a difference only where it survives both. Two composition differences hold under every reportability definition, tow definition and weighting of the human side, and under both definitions of driverless operation: turn-across-path crashes are under-represented among ADS crashes (0.23--0.46$\times$ the CRSS share under the primary driverless definition) and so are single-vehicle crashes (0.19--0.56$\times$). In severity-floor rear-end crashes the ADS vehicle is the striking party at a quasi-induced-exposure ratio of 0.121 (95\% CI [0.05, 0.23]) against a human baseline of 1.000. We also document three pitfalls in the federal data: an operator-specific coding of driverless operation, a tow-away criterion confounded by fleet recovery practice, and an occupancy confound that accounts for about 42\% of the apparent ADS--human severity gap.
\end{abstract}


\section{Introduction}
\label{sec:intro}

Evidence on the safety of automated driving systems (ADS) has so far centered on crash rates: crashes per mile for an ADS fleet compared with human drivers in the same areas. The most recent such analysis, by \citet{teoh2026}, finds that Waymo vehicles in driverless operation had police-reportable crash involvement rates 68\% lower than human drivers in the same areas and years. Rate comparisons of this kind depend on an exposure denominator, which in practice is available for only a few operators.

This paper asks a complementary question that requires no denominator. Given that a crash has occurred, does the composition of crash types differ between ADS and human-driven vehicles? A composition comparison cannot say whether ADS vehicles crash more or less often, but it can show where their crashes are structurally different from human ones, and it can include every ADS operator that files under the NHTSA Standing General Order \citep{nhtsa_sgo}.

Making the two sides comparable is the main difficulty. The ADS side is the SGO incident files, which are filed by operators, carry free-text narratives, and compel reporting far below the police-reporting threshold. The human side is the Crash Report Sampling System \citep{nhtsa_crss}, a nationally representative sample of police-reported crashes. We harmonize them in three ways: a CRSS-compatible extraction layer that codes each SGO narrative into the federal crash-type taxonomy, building on an earlier LLM extraction pipeline for ADS crash narratives \citep{jang2026extraction}; two bracketing definitions of police reportability; and a restriction of both sides to a shared window and operating domain.

The contributions are as follows.
\begin{itemize}
  \item A documented assembly of the SGO ADS corpus across both file generations, including deduplication and two definitions of driverless operation (Sections~\ref{sec:scope}--\ref{sec:assembly}).
  \item Composition differences that hold under every reportability definition, all three tow definitions, reweighting of the human side onto the ADS domain and both driverless definitions, with checks on how much LLM coding error could explain them (Sections~\ref{sec:survives}--\ref{sec:misclass}).
  \item A quasi-induced-exposure estimate of the ADS striking propensity in rear-end crashes, with a conditional-sequence analysis of how the ADS vehicle is moving at impact (Section~\ref{sec:qie}).
  \item Three data pitfalls with quantified effects: the operator-specific coding of driverless operation, the tow confound, and the occupancy confound (Sections~\ref{sec:assembly}, \ref{sec:tow} and~\ref{sec:occupancy}).
  \item Accuracy and proxy validation of the extraction layer, including a lower bound on how often narratives omit facts recorded in the structured fields (Sections~\ref{sec:extraction}--\ref{sec:proxy}).
\end{itemize}

\section{Related work}
\label{sec:related}

\paragraph{Crash rates against human benchmarks.} \citet{kusano2024} compared Waymo rider-only crashes reported under the SGO over 7.1 million miles with human benchmarks from the literature, finding an 80\% lower any-injury-reported and a 55\% lower police-reported crashed-vehicle rate. \citet{kusano2025} extended the comparison to 56.7 million miles and disaggregated it into eleven crash-type groups: the largest reduction was in vehicle-to-vehicle intersection crashes, and single-vehicle crashes were among the types significantly reduced for any-injury-reported outcomes. \citet{dilillo2024} used third-party liability insurance claims against a zip-code-calibrated human benchmark and also found substantial reductions. With \citet{teoh2026}, these studies need operator mileage, which confines them to the operators that publish it. This paper needs no mileage and covers every SGO filer, at the cost of saying nothing about rates. The two approaches point the same way: a crash type whose rate falls by more than the total falls as a share of crashes, so the intersection and single-vehicle reductions in \citet{kusano2025} are consistent in direction with the composition differences reported here.

\paragraph{Crash characteristics.} \citet{favaro2017} analysed the early California DMV reports from AV testing (September 2014 to March 2017) and found that most were rear-end collisions in which the AV was struck from behind, usually at or near zero speed. The quasi-induced-exposure analysis in Section~\ref{sec:qie} quantifies that pattern against a human baseline. The closest prior work is \citet{abdelaty2024}, a matched case-control study of 2,100 ADS and ADAS vehicle accidents, drawn from California DMV reports and SGO filings, against 35,113 human-driven vehicle accidents from SWITRS. They find ADS accidents less likely than human-driven ones in most matched scenarios, but 5.25 times more frequent under dawn/dusk conditions and 1.98 times under turning conditions. Their turning result appears to run against the turn-across-path finding here, but the two designs answer different questions. They estimate the relative odds of an accident within scenarios matched on conditions, turning among them, with ADS and ADAS vehicles both in the data. We compare the mix of crash types given that a crash occurred, restrict the ADS side to driverless operation and to 2021--2024, bracket police reportability on both sides, and separate turn-across-path from the other turning configurations. The two results are not directly comparable, and we do not attempt to reconcile them at the case level.

\section{Scope}
\label{sec:scope}

The corpus is the NHTSA Standing General Order (SGO) ADS incident files, both file generations, deduplicated to distinct crashes. The current-generation file is the release that runs to incident month August 2026. The human comparator is CRSS at the vehicle level, restricted to the ADS operating domain. A CRSS-compatible extraction layer codes each narrative into the same crash-type taxonomy the federal files use.

The full corpus is all 2,450 in-scope incidents, spanning 2021--2026. The comparison window is the 828 of those with an incident year in 2021--2024, the CRSS releases that exist; every ADS-versus-human comparison runs on this. The driverless window is the 456 or 576 of those flagged driverless, depending on which definition is used (Section~\ref{sec:assembly}).

\section{Corpus assembly}
\label{sec:assembly}

\paragraph{Deduplication.} Same Incident ID is commonly described as entity-scoped. It is not: it already links 149 Cruise/GM and 139 Waymo/Transdev groups. The residual after it is 5 pairs in 2,668 incidents (0.19\%), all five confirmed by inspection---their narratives cross-reference each other's report identifiers. Table~\ref{tab:funnel} gives the full funnel.

\paragraph{Driverless} is defined two ways, because operators file differently. Waymo files \fld{None} for \fld{Driver / Operator Type} for driverless operation; Cruise never does, filing \fld{Remote (Commercial / Test)}. Treating \fld{None} as the only driverless value does not select driverless crashes---it selects Waymo crashes (Table~\ref{tab:driverless}).

\begin{table}[t]
\centering
\small
\begin{tabular}{@{}lrl@{}}
\toprule
Step & $n$ & Note \\
\midrule
Raw ADS filings (both generations) & 3,773 & \\
One row per Report ID & 3,336 & highest Report Version retained \\
ADS verified engaged & 3,253 & $-83$ \\
Public road & 3,031 & $-222$ parking lot \\
Distinct incidents & 2,668 & Same Incident ID + 5 verified merges \\
\midrule
\multicolumn{3}{@{}l}{\emph{Two branches from 2,668, neither nested in the other:}} \\
\quad Extraction-evaluation frame & 2,653 & $-15$ narrative too short \\
\quad Full corpus (after narrative exclusions) & 2,450 & $-218$; see caption \\
\midrule
\multicolumn{3}{@{}l}{\emph{Restricted to the comparison window, 2021--2024:}} \\
\quad Comparison window, all supervision & 828 & denominator for tier counts below \\
\qquad Driverless, \defn{strict} & 456 & \\
\qquad Driverless, \defn{no onboard} & 576 & \\
\bottomrule
\end{tabular}
\caption{ADS corpus funnel. Narrative exclusions are IIHS's remaining criteria, which have no structured counterpart in SGO and are therefore extracted; only an affirmative \emph{no} excludes, so unknown and missing-extraction records are retained and counted. The corpus runs to 2026 while CRSS ends at 2024, so roughly two-thirds of incidents (1,622 of 2,450) fall outside the comparison window---which is why the window denominator is stated wherever a share is reported. Both branches descend from 2,668, not from each other. Extraction accuracy (Table~\ref{tab:extraction}) is scored on the 2,653 incidents with a usable extraction, before the narrative exclusions, because those exclusions are derived from the extraction's own output; 2,435 of the 2,450 analysis incidents carry an extraction, the other 15 retained under the unknown-is-not-excluded rule. The three exclusion criteria overlap and do not sum to the total: 38 not a crash, 11 private property and 174 ADS-not-engaged total 223, but 5 filings fail more than one test (4 are both not-a-crash and not-engaged, 1 is both private and not-engaged), so the union removed is 218 and $2{,}668 - 218 = 2{,}450$.}
\label{tab:funnel}
\end{table}

\begin{table}[t]
\centering
\small
\begin{tabular}{@{}lrrrr@{}}
\toprule
Definition & $n$ (window) & Waymo share & non-Waymo & Tier A non-Waymo \\
\midrule
\defn{strict} (\fld{None} only) & 456 & 99.8\% & 1 & 1 \\
\defn{no onboard} (\fld{None} or remote) & 576 & 79.0\% & 121 & 50 \\
\bottomrule
\end{tabular}
\caption{Within the comparison window, the \fld{None}-only rule yields 455 Waymo incidents and one other. Both definitions are carried, because a remote operator may be actively teleoperating, which is arguably a human driving.}
\label{tab:driverless}
\end{table}

\section{Reportability}
\label{sec:reportability}

SGO compels filings far below the police-reporting bar, so the ADS side must be cut down before comparison. Two definitions are bracketed and a difference is claimed only where it survives both: Tier~A, an objective severity floor (injury B or worse, a tow, or an airbag deployment), and Tier~B, a written reasonable-person rubric for police reportability graded \emph{yes}, \emph{maybe} or \emph{no} (Table~\ref{tab:tiers}).

\paragraph{Tier B validation.} Against IIHS's published marginals on their restricted frame \citep{teoh2026}, the rubric yields 26.8\% yes-or-maybe against their 21.6\%. The police-involvement gradient is monotone at 0.133/0.238/0.497 for no / maybe / yes against their 0.16/0.42/0.68. Rubric-compliance error, measured against structured triggers the rubric names in writing, is under 1\%. 

\begin{table}[t]
\centering
\small
\begin{tabular}{@{}llrrrr@{}}
\toprule
 & & \multicolumn{2}{c}{Full corpus} & \multicolumn{2}{c}{Comparison window} \\
\cmidrule(lr){3-4}\cmidrule(l){5-6}
Tier & Definition & $n$ & share & $n$ & share \\
\midrule
A (\defn{any tow}) & injury B+ / any tow / airbag & 1,052 & 42.9\% & 243 & 29.3\% \\
A (\defn{cp tow}) & injury B+ / partner tow / airbag & 384 & 15.7\% & 119 & 14.4\% \\
A (\defn{no tow}) & injury B+ / airbag & 319 & 13.0\% & 103 & 12.4\% \\
B (yes) & reasonable-person rubric & 475 & 19.4\% & 230 & 27.8\% \\
B (yes+maybe) & rubric, pooled & 626 & 25.6\% & 295 & 35.6\% \\
\bottomrule
\end{tabular}
\caption{Reportability tiers under both denominators. The three Tier~A variants exist because the tow leg is confounded on the ADS side (Section~\ref{sec:tow}); Tier~A is only as strong as its stability across them.}
\label{tab:tiers}
\end{table}

\section{The tow confound}
\label{sec:tow}

In the full corpus, SGO records 791 crashes in which only the subject vehicle was towed, against 159 in which the crash partner was towed. Damage does not produce this asymmetry: the ADS vehicle is disproportionately the stationary, struck party, so if tows tracked damage alone the ADS vehicle would be towed less, not more. Fleet operators recover a vehicle that cannot resume autonomous operation regardless of damage, so an any-tow criterion conflates fleet logistics with crash severity.

This confound is why every severity-floor result is reported under all three tow definitions (Table~\ref{tab:tiers}). Under the permissive \defn{any tow} definition, backing and sideswipe same direction appear elevated (10.3$\times$ and 2.2$\times$ respectively). Neither survives the \defn{cp tow} (partner-tow-only) definition: backing empties entirely, since every qualifying crash entered the severity floor solely through an ADS tow, and sideswipe same direction reverses sign to 0.34$\times$.

\section{Findings}
\label{sec:findings}

All results below are on the comparison window (828 incidents), driverless under \defn{no onboard} unless stated.

\subsection{What survives every specification}
\label{sec:survives}

\begin{table}[t]
\centering
\small
\begin{tabular}{@{}lcccc@{}}
\toprule
 & \multicolumn{2}{c}{Ratio to CRSS share, across arms} & \multicolumn{2}{c}{Multinomial} \\
\cmidrule(lr){2-3}\cmidrule(l){4-5}
Category & \defn{no onboard} & \defn{strict} & OR & 95\% CI \\
\midrule
turn across path & 0.23--0.46 & 0.09--0.32 & 0.40 & [0.25, 0.75] \\
single vehicle & 0.19--0.56 & 0.16--0.42 & 0.16 & [0.09, 0.31] \\
\bottomrule
\end{tabular}
\caption{The two composition findings that hold in every specification. For each driverless definition, the range is the minimum and maximum ADS-to-human share ratio across six specifications: the five reportability definitions of Table~\ref{tab:tiers} (Tier~A under each of its three tow definitions, and Tier~B yes and yes+maybe) with unweighted CRSS, plus Tier~A with CRSS entropy-balanced onto the ADS domain. In all twelve specifications the ADS-side 95\% bootstrap interval lies below parity. The multinomial odds ratios are for Tier~A under \defn{no onboard}, conditioned on environment (intersection, lighting, speed); their intervals resample PSUs within (year, stratum) on the human side and incidents on the ADS side. Under \defn{strict} the point estimates are 0.37 and 0.15.}
\label{tab:survives}
\end{table}

Table~\ref{tab:survives} reports the two composition differences that hold in every specification: turn-across-path and single-vehicle crashes are both markedly under-represented among ADS crashes relative to human crashes, under both driverless definitions.

\subsection{Robustness to crash-type coding error}
\label{sec:misclass}

Both findings rest on the extracted crash type, whose overall accuracy against the structured key is 0.744 (Table~\ref{tab:extraction}). The overall figure is the wrong quantity for either finding, so each is checked separately.

\paragraph{Single vehicle.} The structured key identifies single-vehicle crashes directly, from a crash partner coded as a fixed object or a pole or tree. Against it the extraction has precision 0.843 and recall 1.000 (59 keyed single-vehicle crashes in the extraction-evaluation frame): it over-calls the category rather than missing it. Correcting the ADS share for that error (multiplying by precision over recall) lowers every ratio, to 0.16--0.48 under \defn{no onboard} and 0.14--0.35 under \defn{strict}. Extraction error therefore works against this finding, not for it.

\paragraph{Turn across path.} The structured key cannot score this category. Contact geometry does not say which way the vehicles were heading, so the per-class accuracy of turn-across-path coding is not measurable from SGO. Two checks bound its effect instead. First, if the extractor never assigns turn-across-path to a crash of another type, the observed ratio equals its recall times the true ratio. Closing the gap to parity would then require a recall no higher than the observed ratio itself---0.23 to 0.46 under \defn{no onboard}---so the extractor would have to miss more than half of all turn-across-path crashes. Second, a turn-across-path crash requires one vehicle to be turning, so SGO's structured pre-crash movement gives an extraction-free upper bound. Counting every crash in which either vehicle is coded as making a left turn or a U-turn against a motor-vehicle partner, as though all were turn-across-path, the ratio is still 0.57--0.76 under \defn{no onboard} and 0.67--0.82 under \defn{strict} for the five unweighted specifications. The bound does not hold everywhere. Under entropy-balanced CRSS it exceeds parity (1.24 and 1.35). If every crash whose movement is coded \fld{Other, see Narrative} or unknown is counted as well, it exceeds parity in nine of the ten unweighted specifications. The turn-across-path finding is therefore robust to coding error against unweighted CRSS, but not established independently of the extraction in the balanced arm.

\subsection{Quasi-induced exposure}
\label{sec:qie}

Conditional on a rear-end crash clearing the objective severity floor, the ADS vehicle is the striking party at a ratio of 0.121 (95\% CI [0.05, 0.23]) against a human baseline of exactly 1.000 estimated from CRSS \citep[on the quasi-induced-exposure method, see][]{stamatiadis1997}. The result holds at 0.077 to 0.150 across reportability definitions, at 0.114 ([0.02, 0.23]) under the \defn{strict} driverless definition, and at 0.143 ([0.00, 0.45]) with the dominant operator excluded.

The conditional-sequence analysis shows how the ADS vehicle is moving at impact: within junction and lateral crash types the ADS vehicle is stationary at impact far more often than a human vehicle in the same structural position---0.200 versus 0.004 in intersecting-straight-paths, 0.647 versus 0.063 in same-direction sideswipes, 0.417 versus 0.054 in turn-into-path. In rear-end crashes the pattern is absent (0.521 versus 0.646), as expected: being stopped is already the ordinary way to be rear-ended.

Quasi-induced exposure assumes that involvement as the struck party is proportional to exposure. If the ADS vehicle’s driving style—abrupt braking, for example—raises its rate of being struck, the ratio understates its striking propensity and overstates the ADS–human difference. The stationary-at-impact pattern above is absent in rear-end crashes, so the data do not show this bias operating there.

\subsection{Occupancy}
\label{sec:occupancy}

SGO files no occupancy field, but \fld{Were All Passengers Belted?} carries an explicit \fld{No Passenger In Vehicle} value in both generations, giving 99.8\% coverage of the full corpus. Driverless crashes are 52.1\% unoccupied against 2.9\% of supervised ones---a sanity check on the field, and an independent corroboration of the IIHS estimate of roughly 48\% \citep{teoh2026}.

\begin{table}[t]
\centering
\small
\begin{tabular}{@{}lrrrr@{}}
\toprule
 & ADS all & ADS unoccupied & ADS occupied & Human \\
\midrule
Any-injury share & 0.281 & 0.138 & 0.438 & 0.656 \\
$n$ & 167 & 87 & 80 & 103,096 \\
\bottomrule
\end{tabular}
\caption{Occupancy accounts for about 42\% of the apparent severity gap (0.375 raw, 0.218 occupied-only). IIHS identified this confound and treated it as unadjustable.}
\label{tab:occupancy}
\end{table}

The adjustment introduces a selection effect in exchange: an occupied ADS vehicle is on a revenue trip, which may differ in route, time and speed from a repositioning trip. Both arms are reported (Table~\ref{tab:occupancy}).

\section{The extraction layer}
\label{sec:extraction}

\paragraph{Model and settings.} The extraction uses Anthropic's Claude Sonnet 4.6 (API model identifier \texttt{claude-sonnet-4-6}) through the Messages API. Each call sends a fixed system prompt and one narrative, together with the reporting entity's name and the report ID. Appendix A gives the schema, the full prompt and the Tier~B rubric it contains. 

\begin{table}[t]
\centering
\small
\begin{tabular}{@{}lrrr@{}}
\toprule
Field & $n$ scored & Coverage & Accuracy \\
\midrule
\var{on\_public\_road} & 2,607 & 98.3\% & 0.996 \\
\var{kabco\_severity} & 623 & 23.5\% & 0.979 \\
\var{tow\_away\_due\_to\_damage} & 659 & 24.8\% & 0.956 \\
\var{airbag\_deployed} & 1,050 & 39.6\% & 0.953 \\
\var{ads\_engaged\_at\_impact} & 2,605 & 98.2\% & 0.933 \\
\var{striking\_role} & 1,604 & 60.5\% & 0.867 \\
\var{lighting} & 1,050 & 39.6\% & 0.856 \\
\var{sv\_pre\_crash\_movement} & 2,583 & 97.4\% & 0.830 \\
\var{acc\_type\_category} & 1,685 & 63.5\% & 0.744 \\
\bottomrule
\end{tabular}
\caption{Accuracy against the SGO structured key, which the model never sees. Denominator: the 2,653 extraction-evaluation records (Table~\ref{tab:funnel}). Coverage is $n$ scored over 2,653; accuracy is over the scored records only. All three legs of the objective severity floor clear 0.95.}
\label{tab:extraction}
\end{table}

\paragraph{The extracted role inherits operator framing.} Of 26 crashes the contact-area key calls the ADS striking, the model agrees on only 7---27\% recall on the minority class, with the marginal striking share falling from 10.7\% to 4.1\%. The narratives are written by the operator, and even instructed to use contact geometry alone the model drifts toward the text's framing. Quasi-induced exposure therefore runs on the structured key, with the extracted layer as a sensitivity that moves the result further in the same direction.

\section{Validation by proxy}
\label{sec:proxy}

\paragraph{Tier~A sufficiency.} The rubric names injury and airbag deployment as triggers in writing, and SGO records both structurally. A crash with a structured injury or airbag deployment that the rubric labels \emph{no} is therefore a candidate failure, countable without adjudication (Table~\ref{tab:proxy}, labelled-no column). Naming of a police agency is counted the same way.

\paragraph{Metamorphic tests.} The first two checks catch only false negatives, since nothing in SGO asserts that a crash was not reportable. Metamorphic tests cover the other direction. Twelve controlled edits, each tied to a rubric clause and with a known correct direction, were applied to 50 narratives, giving 533 cases, of which 440 are informative. Edits adding a trigger (injury, police attendance, a partner tow, an airbag, a hit and run) raised the label in 94--100\% of cases. Edits the rubric says to ignore left the label unchanged in only 69--86\% of cases, 82\% overall: sensor-only damage 69\%, a bare statement of damage 85\%, NHTSA-filing boilerplate 85\%, an AV tow 85\%, renaming the operator 86\%, and removing a tow 81\%. Replacing described damage with a statement that neither vehicle was damaged lowered the label in 13 of 24 cases. The rubric thus applies its named triggers reliably but moves under irrelevant text in about 18\% of cases. 

\begin{table}[t]
\centering
\small
\begin{tabular}{@{}lrrrr@{}}
\toprule
Structured trigger & $n$ & labelled no & narrative mentions it & silent \\
\midrule
Injury (minor+) & 241 & 20 (8.3\%) & 2 (0.8\%) & 18 (7.5\%) \\
Airbag deployed & 112 & 20 (17.9\%) & 0 (0.0\%) & 20 (17.9\%) \\
Police agency named & 512 & 240 (46.9\%) & 3 (0.6\%) & 237 (46.3\%) \\
\bottomrule
\end{tabular}
\caption{Apparent failures decompose into rubric error (under 1\%) and the narrative omitting a fact the structured field records. Percentages are of $n$; the last two columns sum to the labelled-no column.}
\label{tab:proxy}
\end{table}

This converts the reportable share from a point estimate into a lower bound with a quantified gap: 7.5\% of injury crashes, 17.9\% of airbag crashes and 46.3\% of police-involved crashes are labelled not reportable because their narratives never mention the trigger (Table~\ref{tab:proxy}, silent column).

\section{Limitations}
\label{sec:limitations}

\begin{itemize}
  \item \emph{Composition, not rate.} The design needs no exposure denominator, and for the same reason it says nothing about whether ADS vehicles crash more or less often than human drivers. It complements rate studies such as \citet{teoh2026} rather than substituting for them.
  \item \emph{Reporting thresholds.} SGO and police reporting differ in what they capture. The two reportability tiers bracket that difference; they do not remove it, and the proxy validation in Section~\ref{sec:proxy} shows the reportable share is a lower bound.
  \item \emph{Operator-authored narratives.} The narratives are written by the operators, and the extracted striking role inherits their framing (Section~\ref{sec:extraction}). Quasi-induced exposure therefore uses the structured key.
\end{itemize}

\section{Conclusion}
\label{sec:conclusion}

A harmonized comparison of SGO ADS crashes against CRSS finds two composition differences that hold under every reportability definition, all three tow definitions, reweighting of the human side onto the ADS domain and both driverless definitions: ADS crashes contain far fewer turn-across-path and single-vehicle crashes than human crashes. In rear-end crashes the ADS vehicle is rarely the striking party. Several apparent differences do not survive: backing and same-direction sideswipes look elevated only under a tow criterion that is confounded by fleet recovery, and roughly 42\% of the apparent severity gap is explained by unoccupied ADS vehicles. Analyses of the federal ADS files should carry both driverless definitions, all three tow definitions, and an explicit statement of the comparison-window denominator.

\section*{Data availability}
All inputs are public: the NHTSA Standing General Order incident report files \citep{nhtsa_sgo} and the NHTSA Crash Report Sampling System \citep{nhtsa_crss}.

\appendix
\section{Extraction schema and prompt}
\label{app:extraction}

\paragraph{Schema.} Every call must return one object with the fields below; the enumerated values are enforced by the tool schema.

\begin{small}
\begin{longtable}{@{}p{0.30\textwidth}p{0.66\textwidth}@{}}
\toprule
Field & Allowed values \\
\midrule
\endhead
\var{acc\_type\_category} & rear\_end, sideswipe\_same\_direction, head\_on, sideswipe\_opposite\_direction, turn\_across\_path, turn\_into\_path, straight\_paths\_intersecting, backing, single\_vehicle, pedestrian\_animal, other\_unknown \\
\var{striking\_role} & striking, struck, undetermined, not\_applicable \\
\var{role\_evidence} & free text \\
\var{sv\_pre\_crash\_movement}, \var{cp\_pre\_crash\_movement} & going\_straight, stopped\_in\_roadway, decelerating, accelerating, starting\_in\_road, turning\_left, turning\_right, making\_u\_turn, changing\_lanes, merging, passing, backing, negotiating\_curve, parked, entering\_parking, leaving\_parking, other, unknown \\
\var{relation\_to\_junction} & intersection, intersection\_related, driveway\_access, non\_junction, unknown \\
\var{intersection\_type} & four\_way, t\_intersection, y\_intersection, roundabout\_traffic\_circle, five\_or\_more, not\_an\_intersection, unknown \\
\var{kabco\_severity} & K, A, B, C, O, unknown \\
\var{tow\_away\_due\_to\_damage}, \var{airbag\_deployed} & yes, no, unknown \\
\var{damage\_descriptor} & none, scratch\_scuff, panel\_dent, deformation, intrusion, sensor\_only, unknown \\
\var{sensor\_only\_damage} & true, false \\
\var{posted\_speed\_limit\_bin} & le\_25, mph\_30\_35, mph\_40\_45, gt\_45, unknown \\
\var{lighting} & daylight, dark\_lighted, dark\_unlighted, dawn\_dusk, unknown \\
\var{police\_reportable} & yes, maybe, no \\
\var{reportable\_evidence} & free text (verbatim span) \\
\var{reportable\_confidence}, \var{confidence} & number in [0, 1] \\
\var{is\_true\_crash}, \var{on\_public\_road}, \var{ads\_engaged\_at\_impact} & yes, no, unknown \\
\bottomrule
\end{longtable}
\end{small}

\paragraph{System prompt.} Sent verbatim on every call. The Tier~B rubric is the section beginning ``REPORTABILITY''.

\begin{lstlisting}
You are a careful traffic-safety analyst coding autonomous-vehicle (AV) collision narratives into the structured fields NHTSA uses for police-reported crashes (the CRSS coding scheme). You receive one narrative at a time and return only the structured fields.

GENERAL RULES
- Code ONLY what the narrative supports. Do not infer beyond the text. Where the narrative is silent, use "unknown" (or the relevant not-applicable value). Abstaining is correct behaviour and is scored as such; guessing is not.
- "subject vehicle" (SV) = the reporting entity's AV. "crash partner" (CP) = the other road user.

CRASH TYPE (acc_type_category) -- classify the FIRST harmful event:
- rear_end: front of one vehicle into the rear of another, same direction.
- sideswipe_same_direction: lateral contact, both travelling the same way.
- head_on: front-to-front, opposite directions.
- sideswipe_opposite_direction: lateral contact, opposite directions.
- turn_across_path: one vehicle turns ACROSS the other's path (the classic left turn across oncoming traffic).
- turn_into_path: one vehicle turns INTO the path of another already travelling on the road it turns onto.
- straight_paths_intersecting: both going straight on crossing roads -- the intersection T-bone.
- backing: either vehicle reversing.
- single_vehicle: no second road user; fixed object, road departure.
- pedestrian_animal: a pedestrian, cyclist, scooter rider or animal struck.
- other_unknown: none of the above, or the narrative does not say.

STRIKING ROLE (striking_role) -- of the SUBJECT vehicle, from CONTACT GEOMETRY, never from fault:
- striking: the SV's front (or, when reversing, its rear) made the contact, or the SV was the vehicle that moved into the other's path.
- struck: the other vehicle made the contact with a stationary or lawfully-proceeding SV.
- undetermined: both moved into each other, or the narrative does not say which.
- not_applicable: no second motor vehicle.
Do NOT reason about who was at fault, who was careless, or who had right of way. A vehicle stopped at a light that is rear-ended is "struck" even if the narrative blames it for stopping abruptly.

MODE AT IMPACT (sv_pre_crash_movement, cp_pre_crash_movement): what each vehicle was doing immediately before the first harmful event.

SEVERITY (kabco_severity): the KABCO grade a police officer would assign to the most severely injured person -- K fatal, A suspected serious, B suspected minor, C possible injury or complaint of pain, O no apparent injury. Use "unknown" when the narrative does not describe injuries at all; do NOT default to O.

DAMAGE (damage_descriptor, sensor_only_damage): describe the worst damage to ANY vehicle. Set sensor_only_damage true only when the ONLY damage described is to exposed sensing hardware -- lidar pod, radar, camera housing, a sensor-bearing mirror unit -- with no damage to structure, panels or glazing.

OPERATING DOMAIN (posted_speed_limit_bin, lighting): code these only if the narrative states or clearly implies them. "unknown" is the common and correct answer.

EXCLUSION FLAGS -- each is yes / no / unknown. "unknown" is a real and correct answer when the narrative does not say; do not guess, because these decide which crashes enter the analysis at all.
- is_true_crash: "no" when the narrative describes something that was not a collision at all -- a suspected contact that proved to be road debris or a pothole, a filing made out of caution with no contact described, a third party opening a door against a stationary AV.
- on_public_road: "no" for private property, parking structures, depots. "unknown" when the narrative does not locate the crash.
- ads_engaged_at_impact: was the automated driving system engaged at the moment of the first harmful event.

REPORTABILITY. Judge: would a reasonable person involved in this crash have reported it to the police?

START FROM "no" AND MOVE UP ONLY ON POSITIVE EVIDENCE. Most contacts between vehicles that then drive away are never reported to anyone. "No" is the default and the most common correct answer; it is not a failure to find something.

Answer "yes" ONLY when the narrative states one or more of these, and quote the words that establish it:
  (a) any injury, complaint of pain, ambulance, or hospital transport;
  (b) an airbag deployed in any vehicle;
  (c) a vehicle was towed BECAUSE it could not be driven. For the CRASH PARTNER, a stated tow is enough. For the SUBJECT AV it is not: the narrative must separately establish the vehicle was disabled or undriveable (see the note on AV tows below);
  (d) a vulnerable road user -- pedestrian, cyclist, scooter rider -- was struck;
  (e) a party left the scene without exchanging information (hit and run);
  (f) police attended, an officer responded, or a police report was made;
  (g) someone else's property was damaged and its owner was not present -- a parked car, pole, fence, or building;
  (h) damage described in terms that establish real extent: deformation, crumpling, intrusion, a panel or door that no longer functions, broken glass, a wheel or axle displaced, fluid leaking, a vehicle undriveable.

Answer "maybe" ONLY when the narrative describes damage in terms that could genuinely fall on either side of (h) -- for example "significant damage" with no further detail, or a described impact severe enough to raise the question while the outcome is left unstated. "Maybe" is a narrow category for real ambiguity, not a resting place for anything unclear.

Answer "no" in every other case, including when the narrative describes contact and damage without establishing extent.

FIVE THINGS THAT ARE NOT EVIDENCE OF REPORTABILITY. Do not let any of them move your answer up:

  1. THE SUBJECT AV BEING TOWED, ON ITS OWN. Fleet operators recover an automated vehicle that cannot continue autonomously -- after a sensor is knocked out of calibration, after a safety-system fault, or simply because policy forbids resuming the trip. So "the AV was towed from the scene" does NOT establish that the AV was damaged enough to be undriveable, and it is the single most over-read signal in these narratives. It counts only if the narrative separately says the AV was disabled, undriveable, or damaged in a way that matches trigger (h). A tow of the CRASH PARTNER's vehicle needs no such corroboration: nobody recovers a member of the public's car as fleet policy.

  2. AN UNDIFFERENTIATED STATEMENT OF DAMAGE. "Both vehicles sustained damage", "the Waymo AV sustained damage", "there was damage to both vehicles" -- these are boilerplate. They establish that contact occurred, which you already know, and say nothing about extent. On their own they support "no".

  3. DAMAGE CONFINED TO EXPOSED SENSING HARDWARE -- lidar pods, cameras, radar housings, sensor-bearing mirror units. Expensive to replace, irrelevant to whether a reasonable person calls the police.

  4. THAT THE VEHICLE IS AUTONOMOUS, or which company operates it.

  5. THAT THIS REPORT WAS FILED WITH NHTSA. Every narrative you see was filed. The filing duty is far broader than police reportability and carries no information about it.

reportable_evidence: quote the shortest verbatim span that DECIDES your answer. For "yes", it must be the text establishing the trigger you relied on. For "no", quote the clause that shows the crash was minor, or the strongest damage language present so a reader can see what you judged insufficient.
reportable_confidence: your calibrated probability in [0,1] that a careful human coder applying this same rubric would give your label. Be honest; "no" answers should often be high-confidence.

confidence: your calibrated probability in [0,1] that this WHOLE extraction is correct. This is separate from reportable_confidence. Be honest; do not default to 1.0.

Return the structured object only.
\end{lstlisting}

\paragraph{User message.} Filled with the narrative, the reporting entity and the report ID.

\begin{lstlisting}
Collision narrative (reporting entity: {entity}, report id: {report_id}):
"""
{narrative}
"""

Code this narrative into the CRSS-compatible schema.
\end{lstlisting}

\end{document}